\documentclass[
reprint,
amsmath,amssymb,
aps,
showkeys
]{revtex4-2}

\usepackage{dcolumn}
\usepackage{bm}
\usepackage{url}
\usepackage{hyperref}
\usepackage{qcircuit}
\usepackage{xcolor}
\usepackage{graphicx}
\usepackage{float}
\usepackage{braket}
\usepackage{mathrsfs}
\hypersetup{colorlinks,
allcolors=violet}
\usepackage{enumitem}
\usepackage{tabularx}
\usepackage{color,soul}
\usepackage{wasysym}
\usepackage{xurl}
\usepackage{array}

\begin{document}

\title{A Toffoli Gate Decomposition via Echoed Cross-Resonance Gates}

\author{Muhammad AbuGhanem{$^{1,2}$}}
\address{$^{1}$ Faculty of Science, Ain Shams University, Cairo, 11566, Egypt}
\address{$^{2}$ Zewail City of Science, Technology and Innovation, Giza, 12678, Egypt}

\email{gaa1nem@gmail.com}

\date{\today}

\begin{abstract}

Quantum computing has garnered significant interest for its potential to solve certain computational problems much faster than the best-known classical algorithms. A fully functional and scalable quantum computer could transform various fields such as scientific research, material science, chemistry, and drug discovery. However, in the noisy intermediate-scale quantum (NISQ) era, quantum hardware faces challenges including decoherence, gate infidelity, and restricted qubit connectivity. Efficient implementation of multi-qubit gates is critical to advancing quantum computing, especially given the constraints of near-term quantum hardware, such as the absence of all-to-all qubit connectivity. Among these gates, the Toffoli gate (or CCNOT gate) plays a pivotal role in a wide range of quantum algorithms and error correction schemes. While various decomposition strategies have been proposed, they often assume idealized all-to-all connectivity, which is not available on most NISQ hardware. This paper introduces a novel decomposition of the Toffoli gate using Echoed Cross-Resonance (ECR) gates, a native operation for many superconducting qubit architectures, including IBM Quantum’s hardware. By leveraging the inherent compatibility of ECR gates with superconducting qubit technology, this approach is intended to facilitate the implementation of the Toffoli gate, potentially reducing circuit depth and enhancing the efficiency of quantum circuits implementations on near-term quantum hardware.

\end{abstract}


\maketitle

\section{Introduction}

Quantum computing represents a transformative leap in solving complex problems that are infeasible for classical computers~\citep{NISQ24,Light}. With potential applications spanning artificial intelligence, cryptography, materials science, drug discovery, and communications, quantum technologies are poised to reshape numerous industries~\citep{IBMQuantum}.

Despite rapid advancements in quantum technologies, many of these hardware limitations persist, making the optimization of quantum circuits critical for maximizing the performance of existing devices~\citep{NISQ18}. As quantum computing progresses towards practical applications~\citep{NISQ24}, overcoming the limitations of noisy intermediate-scale quantum (NISQ) devices becomes crucial~\citep{NISQ18}. These devices face challenges such as limited qubit connectivity, gate infidelity, and noise, which complicate the implementation of efficient quantum algorithms and restrict scalability~\citep{scalability}.

The primary challenge in constructing a quantum computer lies in the fragility of quantum information, which is highly susceptible to various noise sources~\citep{GoogleQuantum}. This vulnerability arises because isolating quantum systems from external disturbances conflicts with the ability to control them effectively for computations, making noise an unavoidable issue~\citep{High-threshold}. Noise sources include qubit imperfections, control systems, deficiencies in materials, errors in state preparation and measurement, and external factors like stray electromagnetic fields or cosmic rays~\citep{High-threshold_15}. While some noise can be mitigated through improved controls~\citep{High-threshold_16}, better materials~\citep{High-threshold_17}, and advanced shielding techniques~\citep{High-threshold_18,High-threshold_19,High-threshold_20}, other sources are inherently difficult or even impossible to eliminate~\citep{High-threshold_1,High-threshold_2,High-threshold_3}.

Efficient implementation of multi-qubit gates, form the backbone of quantum algorithms and error correction protocols, enabling their practical execution on quantum hardware~\citep{IBMQuantum,PhotonicQuantumComputers}. Among these, the Toffoli gate—also known as the CCNOT or CCX gate— 
is of particular significance~\citep{Toffoli}. The gate's functionality makes it indispensable for applications such as quantum arithmetic, reversible logic synthesis, Grover's search algorithm, and quantum error correction schemes (see Section~\ref{sec:theory}). However, its multi-qubit nature poses challenges for near-term quantum devices due to limited qubit connectivity, intrinsic noise, and hardware-specific constraints~\citep{NISQ18}.

While various decomposition strategies have been proposed, they often assume idealized all-to-all connectivity, which is not available on most (near-tearm) quantum hardware. This gap highlights the need for hardware-specific optimizations that address practical constraints~\citep{NISQ18}. This work proposes a novel decomposition of the Toffoli gate using Echoed Cross-Resonance (ECR) gates, a native operation in many superconducting qubit systems, including IBM Quantum's quantum hardware~\citep{IBMQuantum}. Unlike traditional approaches, the ECR-based decomposition leverages hardware-native operations to minimize circuit depth and mitigate noise. 

The ECR gate~\citep{cross-resonance, ECR_Hamiltonian, 64QV, ibm433} is a 2-qubit entangling gate native to superconducting qubit architectures, particularly in IBM Quantum's quantum hardware. It operates by driving one qubit (control) at the resonance frequency of another qubit (target), inducing an effective \(ZX\)-type interaction. The “echoed" design involves sandwiching the cross-resonance pulse between two \(\pi\)-pulses, effectively canceling out unwanted 1-qubit \(Z\) errors and enhancing gate fidelity. This operation, when combined with local 1-qubit rotations, can implement universal 2-qubit gates, such as CNOT. Unlike the standard CNOT gate, the ECR gate's implementation on superconducting qubits aligns better with hardware constraints like limited connectivity and reduces the overhead of additional calibration steps.

By leveraging the ECR gate's properties, our approach aims to achieve improved compatibility with superconducting qubit platforms, minimizing circuit depth and addressing connectivity limitations inherent in NISQ devices, paving the way for practical implementation and optimizing performance of Toffoli gates and similar multi-qubit operations on near-term quantum devices.

\section{Qubit Connectivity constraints in the NISQ devices}

Quantum computers are subject to physical constraints based on their hardware architecture, especially when it comes to how the qubits are physically arranged and connected~\citep{NISQ18}. In many quantum processors of the NISQ era, qubits are not uniformly connected to all other qubits, meaning that the ability to interact with one another is restricted~\citep{NISQ18}. As a result, to implement certain quantum operations, information often needs to be moved between qubits that are not directly linked. This requires qubit swaps, which can add extra steps and complexity to quantum circuits.

For quantum devices in the NISQ era~\citep{NISQ18}, these connectivity limitations are especially problematic. The relatively small scale of these systems, combined with noise and error-prone operations, means that a significant portion of a quantum algorithm may be spent simply re-routing qubit information. This often adds both depth and potential error to circuits, as more swaps introduce more chances for things to go wrong. Effectively, these restrictions mean that NISQ devices are often forced to work around their architecture~\citep{MSc}, reducing the efficiency and scalability of quantum algorithms~\citep{scalability}.

One of the most commonly used techniques for qubit entanglement on such devices is the cross-resonance (CR) gate, which enables interaction between pairs of qubits~\citep{cross-resonance, ECR_Hamiltonian, 64QV, ibm433}. The CR gate typically functions like a controlled-$R_x$ gate, and when its rotation angle is adjusted to $\pi$, it can mimic a CNOT gate, provided that the appropriate 1-qubit operations are performed as well. However, due to imperfections and the noise that is characteristic of NISQ devices, implementing the CR gate is not always straightforward~\citep{NISQ18}. To counteract the resulting errors, it’s common practice to use two half-cross-resonance pulses with opposite polarities in sequence. This pulse engineering technique helps to cancel out some of the errors, but it also highlights how current hardware limitations still impose significant challenges on gate fidelity and overall performance~\citep{Hardware-Conscious}.

\begin{figure}
    \centering
    \includegraphics[width=0.47\textwidth]{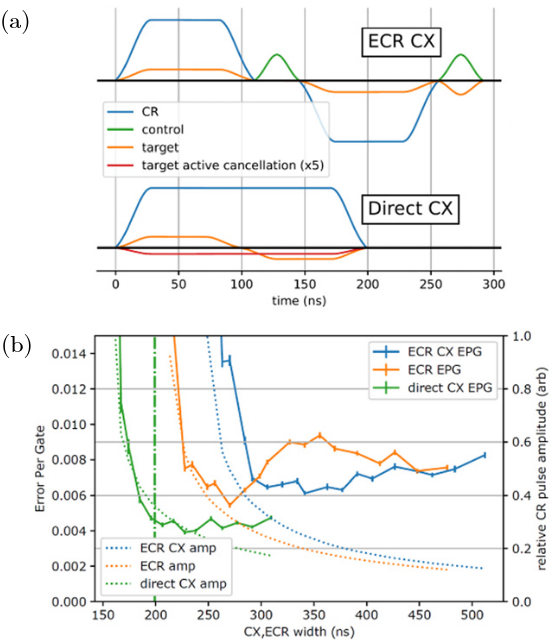}
    \caption{(a) Comparison of pulse profiles for the echoed cross resonance (ECR) CX gate and the direct CX gate implementation on the C$_{22}$-T$_{19}$ pair on \textit{ibmq\_montreal}. (b) Gate error as a function of gate time for the ECR CX (blue), ECR (orange), and direct CNOT (green) gates. The dotted lines show the CR-drive amplitudes for different gate types and durations. The vertical dashed line indicates the gate time used for the direct CX in QV64. Regenerated under Creative Commons Attribution 4.0 license (\url{ https://creativecommons.org/licenses/by/4.0/}) from~\citep{64QV}.}
    \label{fig:ecr}
\end{figure}

To improve the speed of 2-qubit gates, one approach is to integrate pre- and post-1-qubit rotations into the circuit compilation, optimizing the use of the native ECR gate~\citep{64QV}. This adjustment reduces the gate duration to just the entangling portion. Comparisons of the error rates for ECR CX and the optimized ECR gate, measured via 2-qubit randomized benchmarking, show that gate duration directly impacts error rates~\citep{64QV} (see Figure~\ref{fig:ecr}). Further speed improvements can be achieved by replacing the traditional echo pulse sequence with a more efficient echo-free CX pulse sequence~\citep{64QV}. This method, based on advanced rotary pulsing techniques~\citep{ECR_Hamiltonian}, combines active cancellation and target rotary tones. These tones minimize unwanted interactions like crosstalk and reduce errors like $ZZ$ and $ZY$. This reduction in gate duration leads to a noticeable decrease in 2-qubit gate errors, improving overall performance~\citep{64QV}.

The issue of qubit connectivity in NISQ devices limits not only the number of operations that can be performed but also the efficiency with which these operations are executed~\citep{Hardware-Conscious}. Thus, while cross-resonance gates provide a valuable tool for implementing entangling operations, the limitations in qubit connectivity and the need for error-correction strategies highlight the difficulties faced by NISQ devices~\citep{NISQ18}. These constraints demand innovative approaches to quantum circuit design, including the engineering of pulse sequences to overcome hardware limitations and reduce error rates.

\section{The Echoed Cross-Resonance Gate}

The cross-resonance gate has become a key component in the development of 2-qubit entangling operations~\citep{Entangling,SQSCZ1,SQSCZ2} for superconducting quantum computers ~\citep{ref10,ref11,ref12}. It creates entanglement by applying microwave pulses, which eliminates the need for adjustable qubit frequencies or tunable couplings~\citep{ECR_Hamiltonian}. This makes it an attractive option for scaling up quantum systems, as it reduces the complexity of control systems and the number of input lines needed. In comparison to other methods that rely on varying magnetic flux to adjust qubit frequencies~\citep{ref13,ref14,ref15} or coupling strengths~\citep{ref16}, or that use microwave pulses on bus systems to generate entanglement~\citep{ref17}, the CR gate offers a more streamlined and efficient approach.

Although CR pulses effectively boost the entangling interaction between qubits, they also introduce undesirable errors that can hinder the execution of high-fidelity gates~\citep{ref18,ref19,ref20,ref21}. A straightforward solution, such as an echo sequence, can eliminate most of these errors in the CR Hamiltonian, leading to a noticeable improvement in 2-qubit gate fidelity~\citep{ref22}. Additionally, errors can be further minimized by carefully selecting an appropriate calibration frame~\citep{ref23} and implementing strategies to mitigate the effects of classical crosstalk~\citep{ref18}. However, as qubit coherence improves, we encounter more complex errors, such as those caused by persistent $ZZ$ interactions, which ultimately limit gate fidelity even when coherence is high.

\begin{figure*}
    \centering
    \includegraphics[width=0.7\textwidth]{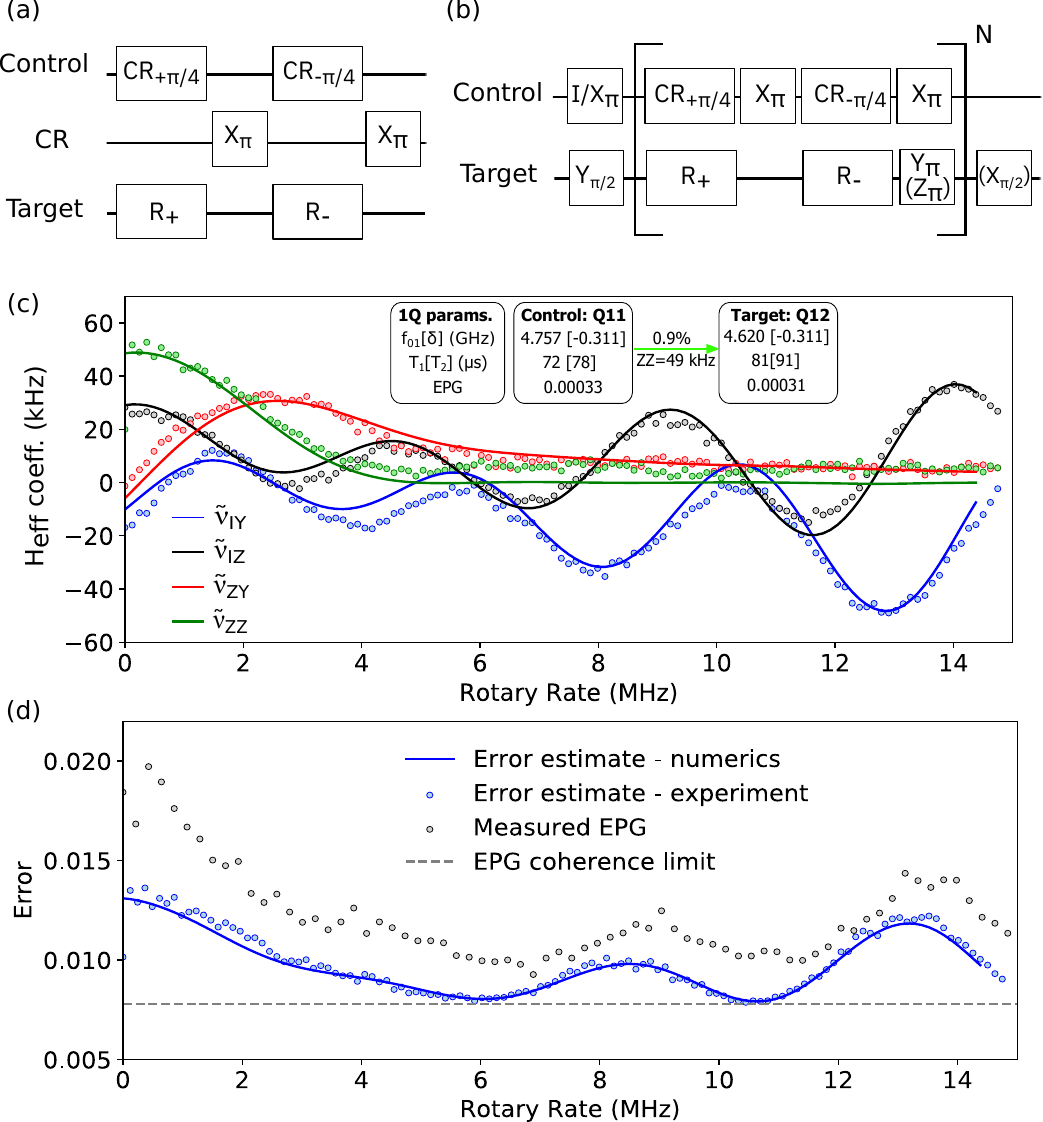}
    \caption{(a) Schematic of the echoed cross resonance (CR) pulse sequence combined with a target rotary pulse, where $R_{\pm}$ represents a fixed $\pm X$ rotation applied to the target qubit.
    (b) Example of (HEAT) Hamiltonian Error Amplifying Tomography, using four distinct gate sequences to accurately determine the Hamiltonian error terms for the echoed CR process.
    (c) Hamiltonian errors for a qubit pair with target rotary pulses, with solid lines representing the numerical fits to the experimental data.
    (d) Two-qubit error per gate (EPG) data (black circles) compared to the theoretical error (blue) calculated by adding the contributions from the four error terms in (c) to the coherence limits set by the measured $T_1$ and $T_2$ values for a gate time of 484 ns (dashed line). Reproduced under the terms of the Creative Commons Attribution 4.0 International license (\url{ https://creativecommons.org/licenses/by/4.0/}) from~\citep{ECR_Hamiltonian}.}
    \label{fig:ecr1}
\end{figure*}

The impact of static $ZZ$ coupling extends beyond simply introducing errors within the 2-qubit subspace. It also leads to unwanted entanglement spreading across the system, affecting other qubits—referred to as ``spectators"~\citep{ECR_Hamiltonian}. Although these $ZZ$-induced errors can be mitigated with more complex pulse sequences, such as higher-order echoes that target spectator-induced errors~\citep{ref24}, or by adding extra 1-qubit rotations to correct errors within the 2-qubit subspace, It is demonstrated that a resonant drive applied to the target qubit can reduce both types of errors simultaneously~\citep{ECR_Hamiltonian}. This method does not require extending the duration or complexity of the 2-qubit gate sequence. The approach, referred to as ``target rotary" pulsing. This ``target rotary" pulsing, shown in Figure~\ref{fig:ecr1}~(a), operates alongside the CR drive and switches sign in the standard 2-pulse echo sequence. Numerical evidence suggests that this approach also helps minimize errors caused by classical crosstalk~\citep{ref18,ref19,ref20}.

The ECR gate~\citep{cross-resonance, ECR_Hamiltonian, 64QV, ibm433} is an advanced modification of the CR gate, developed to address challenges associated with imperfections and unwanted interactions in quantum systems, particularly those arising from detuning or noise. This gate is a composite operation that combines the basic CR gate with corrective elements, often involving ``echo" operations, such as applying the gate with reversed drive amplitudes at different times.

The ECR gate implements the operation \(\frac{1}{\sqrt{2}}(IX - XY)\), where \(I\) is the identity matrix and \(X\) and \(Y\) are Pauli matrices. This echoing mechanism is crucial for enhancing the gate's performance by reducing errors and preserving the fidelity of the quantum state in practical experiments. Current IBM Quantum's superconducting quantum devices~\citep{IBMQuantum} utilize the ECR gate for entanglement operations, rather than the CNOT gate typically used in theoretical quantum circuits~\citep{cross-resonance, ECR_Hamiltonian}. While both gates are designed to generate entanglement, the ECR gate’s tailored mechanism is optimized for the hardware's architecture, providing improved performance and error mitigation in practical quantum computing applications.

\section{Echo Cross-Resonance Model}

The dynamics of the Cross-Resonance gate are captured by the Hamiltonian~\citep{ECR_Hamiltonian}:

\begin{equation}\label{eq:Ham}
    \begin{aligned}
{\cal H}&= \sum_{j=0}^1 \left[\epsilon_j \hat{a}_j^\dagger \hat{a}_j + \frac{\delta_{\text{res},j}}{2}\hat{a}_j^\dagger \hat{a}_j \left(\hat{a}_j^\dagger \hat{a}_j - I\right)\right] 
 \\ &+ E_0 \cos(\omega_{\text{drive}} t + \phi_{\text{drive}})\left(\hat{a}_0^\dagger + \hat{a}_0\right)
 \\&+ \lambda \left(\hat{a}_0^\dagger \hat{a}_1 + \hat{a}_0 \hat{a}_1^\dagger\right),
\end{aligned}
\end{equation}

\noindent
where $\hbar = 1$, $\hat{a}_j$ is the lowering operator for the $j\text{-th}$ transmon qubit, $I$ is the identity operator. The system consists of two transmons: the ``control" and the ``target." Each transmon is modeled as a ``Duffing oscillator," described by $\epsilon_j \hat{a}_j^\dagger \hat{a}_j + (\delta_{\text{res},j}/2)\hat{a}_j^\dagger \hat{a}_j \left(\hat{a}_j^\dagger \hat{a}_j - I\right)$, with $\epsilon_j$ representing the frequency and $\delta_{\text{res},j}$ the anharmonicity. The coupling between the transmons is a Jaynes-Cummings interaction $\lambda \left(\hat{a}_0^\dagger \hat{a}_1 + \hat{a}_0 \hat{a}_1^\dagger\right)$. Additionally, the control transmon is driven by a microwave tone characterized by amplitude $E_0$, frequency $\omega_{\text{drive}}$, and phase $\phi_{\text{drive}}$.

Following the method described in~\cite{ref19}, a time-independent block-diagonal Hamiltonian is derived. The process involves diagonalizing the free Hamiltonian, transforming the drive into the dressed basis, shifting to a rotating frame aligned with the dressed target frequency, applying the rotating-wave approximation (RWA), and block-diagonalizing based on the least action principle. This approach holds as long as the eigenvectors of the intermediate Hamiltonian have minimal overlap between blocks, which is achieved when the transmon frequencies are sufficiently detuned~\citep{ECR_Hamiltonian}.

The final block-diagonal Hamiltonian in the qubit-qubit subspace only includes the identity ($I$) and $Z$ terms on the control qubit:

\begin{equation}\label{Eqn:Hamiltonian}
    \begin{aligned}
{\cal H}(E_0) &= \pounds_{IX} \frac{IX}{2} ++ \pounds_{ZI} \frac{ZI}{2} + \pounds_{IZ} \frac{IZ}{2}  \\
&  + \pounds_{ZZ} \frac{ZZ}{2} + \pounds_{ZX} \frac{ZX}{2},
\end{aligned}
\end{equation}

\noindent
where the coefficients $\pounds_j = \pounds_j(E_0)$ depend on the drive amplitude $E_0$. Notably, there are no $IY$ or $ZY$ terms since the target qubit rotation is confined to the $X$ quadrature. Inverting the drive amplitude sign ($E_0 \to -E_0$) reverses the non-diagonal Pauli coefficients' signs, as diagonal terms are even-order and non-diagonal terms are odd-order functions of $E_0$.

The unwanted terms in $H(E_0)$ introduce errors in implementing the ideal $ZX_{\frac{\pi}{2}}$ gate. To mitigate these errors, the Echoed $ZX_{\frac{\pi}{2}}$ gate is employed:

\begin{equation}
    \begin{aligned}
ZX_{\frac{\pi}{2}} &= XI \cdot ZX_{-\frac{\pi}{4}} \cdot XI \cdot ZX_{\frac{\pi}{4}},
 \end{aligned}
\end{equation}

\noindent
where a general $ZX$ rotation by angle $\varpi$, denoted $ZX_{\varpi}$, is given by:

\begin{equation}
    \begin{aligned}
ZX_{\varpi} &= e^{-i \frac{\varpi}{2} ZX}.
 \end{aligned}
\end{equation}

Using the Hamiltonian in Eq.~\eqref{Eqn:Hamiltonian}, the $ZX_{\frac{\pi}{2}}$ gate is implemented as:

\begin{equation}\label{eq:Upulse}
    \begin{aligned}
\hat{\hat{\mho}} &= XI \cdot e^{-i H(-E_0) t} \cdot XI \cdot e^{-i H(E_0) t},
 \end{aligned}
\end{equation}

\noindent
where $t$ represents the gate duration. Analytical solutions for $e^{-i H(\pm E_0) t}$ yield the unitary operator:

\begin{equation}\label{eq:Echo_unitary}
    \begin{aligned}
\hat{\hat{\mho}} &= \zeta_{II} II + \zeta_{IZ} IZ+ \zeta_{IY} IY  + \zeta_{ZX} ZX,
 \end{aligned}
\end{equation}

\noindent
with coefficients:

\begin{equation}
    \begin{aligned}
\zeta_{II} = & \,
\quad \frac{(\pounds_{IX}^2-\pounds_{IZ}^2)\sin(\zeta t/2)\sin(\xi t/2)}{\zeta \xi}
\\&
+ \frac{(-\pounds_{ZX}^2+\pounds_{ZZ}^2)\sin(\zeta t/2)\sin(\xi t/2)}{\zeta \xi}
\\&
+\frac{\zeta \xi\cos(\zeta t/2)\cos(\xi t/2)}{\zeta \xi}
    \end{aligned}
\end{equation}

\begin{equation}
    \begin{aligned}
\zeta_{IZ} = & \,-i\frac{(\pounds_{IZ}-\pounds_{ZZ})\zeta \cos(\zeta t/2)\sin(\xi t/2)}{\zeta \xi}\\
&-i\frac{\xi(\pounds_{IZ}+\pounds_{ZZ})\sin(\zeta t/2)\cos(\xi t/2)}{\zeta \xi} 
    \end{aligned}
\end{equation}

\begin{equation}
    \begin{aligned}
\zeta_{IY} = & -\frac{2i(\pounds_{IX}\pounds_{IZ}-\pounds_{ZX}\pounds_{ZZ})\sin(\zeta t/2)\sin(\xi t/2)}{\zeta \xi}\\
    \end{aligned}
\end{equation}

\begin{equation}
    \begin{aligned}
\zeta_{ZX} = & \quad i\frac{(\pounds_{IX}-\pounds_{ZX})\zeta \cos(\zeta t/2)\sin(\xi t/2)}{\zeta \xi}\\
& -i\frac{\xi(\pounds_{IX}+\pounds_{ZX})\sin(\zeta t/2)\cos(\xi t/2)}{\zeta \xi} 
    \end{aligned}
\end{equation}

\noindent
and $\zeta$, $\xi$ are given by:
\begin{equation}
    \begin{aligned}
\zeta &= \big((\pounds_{IZ}+\pounds_{ZZ})^2+(\pounds_{IX}+\pounds_{ZX})^2\big)^{\frac{1}{2}},
 \end{aligned}
\end{equation}

\begin{equation}
    \begin{aligned}
\xi &=   \left((\pounds_{IZ}-\pounds_{ZZ})^2+(\pounds_{IX}-\pounds_{ZX})^2\right)^{\frac{1}{2}}.
 \end{aligned}
\end{equation}

The effective echoed Hamiltonian, obtained from Eq.~\eqref{eq:Echo_unitary}, simplifies to include only $IY$, $IZ$, and $ZX$ terms dependent on $\pounds$ and $t$. For the $ZX_{\frac{\pi}{2}}$ gate, further analysis of the unitary confirms that it adheres to the expected operation within a desired tolerance. These calculations underpin the robustness of the echoed gate implementation~\citep{ECR_Hamiltonian}. The echoing technique embedded within the ECR gate is designed to cancel out errors, thereby enhancing the overall fidelity of the operation, especially in the presence of noise and imperfections. Further analysis of the Hamiltonian with crosstalk and rotary, readers are referred to~\citep{ECR_Hamiltonian}.

\section{Implementations of the Toffoli Gate} 

\subsection{Theory}\label{sec:theory}

Multi-qubit Toffoli gates, featuring multiple control qubits and a single target qubit, are essential components in quantum computing due to their broad utility and theoretical significance. As a versatile quantum gate, the Toffoli gate~\citep{Toffoli}, when paired with the Hadamard gate, can facilitate universal quantum communication~\citep{TOFFOLI_52,TOFFOLI_51}. Furthermore, it serves as a key example of reversible logic gates~\citep{TOFFOLI_53}, bridging the divide between classical irreversible computing and quantum processing. The Toffoli gate is also widely used in several important quantum algorithms and protocols, such as Shor’s algorithm~\citep{TOFFOLI_54,TOFFOLI_55}, Grover search algorithms (GSA)~\citep{GSA}, quantum error correction techniques (QECs)~\citep{TOFFOLI_60,TOFFOLI_59,TOFFOLI_61,GoogleQuantum}, and fault-tolerant quantum circuits (FTQCs)~\citep{TOFFOLI_58,TOFFOLI_57}.

The $N$-qubit Toffoli gates are a class of reversible quantum logic gates, consisting of $N-1$ control qubits and one target qubit. The target qubit is flipped if all control qubits are in the state $\ket{1}$~\citep{Nilson}. When $N=2$, the multi-qubit Toffoli gate simplifies to the well-known CNOT gate. Here, we focus on the 3-qubit Toffoli gate, commonly referred to simply as the Toffoli gate and denoted as \(\hat{\mho}_{\text{Toffoli}}\) here, which holds significant importance in quantum information science.

The Toffoli gate is a universal reversible gate that operates on three qubits: two control qubits (\(Q_1, Q_2\)) and one target qubit (\(Q_3\)). The target qubit \(Q_3\) flips its state only when both \(Q_1\) and \(Q_2\) are in the \(|1\rangle\) state. The Toffoli gate's ability to conditionally control a target qubit makes it a vital component in quantum computation. Mathematically, its unitary operation \(\hat{\mho}_{\text{Toffoli}}\) is represented as:

\[
\hat{\mho}_{\text{Toffoli}} = \begin{bmatrix}
1 & 0 & 0 & 0 & 0 & 0 & 0 & 0 \\
0 & 1 & 0 & 0 & 0 & 0 & 0 & 0 \\
0 & 0 & 1 & 0 & 0 & 0 & 0 & 0 \\
0 & 0 & 0 & 1 & 0 & 0 & 0 & 0 \\
0 & 0 & 0 & 0 & 1 & 0 & 0 & 0 \\
0 & 0 & 0 & 0 & 0 & 1 & 0 & 0 \\
0 & 0 & 0 & 0 & 0 & 0 & 0 & 1 \\
0 & 0 & 0 & 0 & 0 & 0 & 1 & 0
\end{bmatrix}.
\]

In this case, the target qubit undergoes a flip operation only when both control qubits are in the $\ket{1}$ state (\(\hat{\mho}_{\text{Toffoli}}\ket{1} \otimes \ket{1}\otimes \ket{0} = \ket{1} \otimes \ket{1}\otimes \ket{1}\), \(\hat{\mho}_{\text{Toffoli}}\ket{1} \otimes \ket{1}\otimes \ket{1} = \ket{1} \otimes \ket{1}\otimes \ket{0}\)), while in all other scenarios, the target qubit remains unchanged. The versatility of Toffoli gates enables their use in simulating any classical logic circuit~\citep{Toffoli}.

Simulating a 3-qubit Toffoli gate typically requires a complex setup of at least five 2-qubit gates~\citep{TOFFOLI_49}, or a minimum of six 2-qubit CNOT gates~\citep{Nilson}, assuming only the use of 2-qubit CNOT gates and 1-qubit operations~\citep{TOFFOLI_48}. This underscores the challenges in emulating the functionality of multi-qubit Toffoli gates with simpler 2-qubit gates, especially when considering scenarios with distributed quantum systems.

\subsection{Existing implementations}

The Toffoli gate has been realized in various forms for superconducting qubit architectures, with one of the most widely recognized implementations being the canonical linear Toffoli gate (as depicted in Eq.~\eqref{fig:toffoli_m}). This particular design is among the earliest and most frequently used circuits for executing the Toffoli gate on linearly connected qubits~\citep{Nilson}.

\begin{equation}
\resizebox{0.5\textwidth}{!}{ 
    \Qcircuit @C=0.7em @R=0.7em @!R {
    & \ctrl{1} & \qw & && 
    & \qw & \gate{T} & \ctrl{1} & \qw & \ctrl{1} & \qw & \ctrl{1} & \qw & \qw & \ctrl{1} & \qw & \qw & \qw & \\
    & \ctrl{1} & \qw &&  \hspace{0.2cm}\equiv \hspace{0.7cm} & 
    & \qw & \gate{T} & \targ & \ctrl{1} & \targ & \ctrl{1} & \targ & \gate{T^\dag} & \ctrl{1} & \targ & \ctrl{1} & \qw & \qw &\\
    &  \targ \qw& \qw &&& 
    & \gate{H} & \gate{T} & \qw & \targ & \gate{T} & \targ & \qw & \gate{T^\dag} & \targ & \gate{T^\dag} & \targ & \gate{H} & \qw &\\
    }
    }
    \label{fig:toffoli_m}
\end{equation}

In practical quantum devices, particularly NISQ superconducting devices, implementing multi-controlled NOT gates is further complicated by limited qubit connectivity. This constraint makes the decomposition into fewer number of gates more challenging~\citep{qubitconnectivity18}. For instance, a three-qubit CCX gate can require a minimum of six CNOT gates in addition to several 1-qubit gates~\citep{Nilson}. 

\begin{equation}
\resizebox{0.49\textwidth}{!}{ 
\Qcircuit @C=0.7em @R=0.7em @!R {
& \ctrl{1} & \qw & && 
\qw & \qw & \qw & \ctrl{2} & \qw & \qw & \qw & \ctrl{2} &  \qw&\ctrl{1} &  \qw  & \qw  &\ctrl{1}  & \gate{T}& \qw\\
& \ctrl{1} & \qw &&  \hspace{0.2cm}\equiv \hspace{0.7cm} & 
\qw & \ctrl{1} &  \qw &\qw &\qw & \ctrl{1} &\qw &\qw &\gate{T^\dag} & \targ & \gate{T^\dag}  & \qw & \targ & \gate{S}& \qw\\
&  \targ \qw& \qw &&& 
\gate{H} & \targ & \gate{T^\dag} & \targ & \gate{T} & \targ &\gate{T^\dag} & \targ & \gate{T}&\gate{H} &  \qw  & \qw & \qw &\qw& \qw\\
}
}
\label{eq:CCX_Nielsen}
\end{equation}

Here $H$ denotes the Hadamard gate, $S = i|1\rangle\langle 1| + |0\rangle\langle 0|$, $T = e^{i\pi/4}|1\rangle\langle 1| + |0\rangle\langle 0|$.

\begin{equation}
\resizebox{0.49\textwidth}{!}{ 
\Qcircuit @C=0.7em @R=0.7em @!R {
& \ctrl{1} & \qw & && 
\qw & \qw & \qw & \ctrl{2} & \qw & \qw & \qw & \ctrl{2} & \ctrl{1} &  \gate{T}  & \qw  &\ctrl{1}  &\qw\\
& \ctrl{1} & \qw &&  \hspace{0.2cm}\equiv \hspace{0.7cm} & 
\qw & \ctrl{1} &  \qw &\qw &\qw & \ctrl{1} & \gate{T} & \qw & \targ & \gate{T^\dag}  & \qw & \targ &\qw\\
&  \targ \qw& \qw &&& 
\gate{H} & \targ & \gate{T^\dag} & \targ & \gate{T} & \targ &\gate{T^\dag} & \targ & \gate{T} &  \gate{H}  & \qw & \qw &\qw\\
}
}
\label{eq:CCX_ibm}
\end{equation}

The traditional decomposition methods often assume a fully connected quantum architecture, which is not always feasible with NISQ devices. Alternative decompositions, such as breaking down the 3-qubit Toffoli gate into a larger number (eight) of CNOT gates, with linear connected qubits, have been explored in~\cite{CCX8CNOTS}.

\begin{equation}
\Qcircuit @C=0.7em @R=0.7em @!R {
& \ctrl{1} & \qw & && \gate{T^\dag} & \ctrl{1} & \qw & \ctrl{1} & \qw & \ctrl{1} & \qw & \qw & \ctrl{1} & \qw & \qw  \\
& \ctrl{1} & \qw &&  \hspace{0.2cm}\equiv \hspace{0.7cm} & \gate{T^\dag} & \targ &  \ctrl{1} & \targ & \ctrl{1} & \targ & \gate{T} & \ctrl{1} & \targ & \ctrl{1} & \qw \\
&  \control \qw& \qw &&& \gate{T^\dag} & \qw & \targ & \gate{T^\dag} & \targ & \gate{T} & \qw & \targ & \gate{T} & \targ & \qw \\
}
\label{eq:CCZ}
\end{equation}

\noindent
Such that, the transformation from CCX to CCZ, can be written as:

\begin{equation}
\Qcircuit @C=0.7em @R=0.7em @!R {
& \ctrl{1} & \qw & && 
\qw &\ctrl{2}& \qw &  \qw \\
& \ctrl{1} & \qw &&  \hspace{0.2cm}\equiv \hspace{0.7cm} & \qw & \ctrl{1} &  \qw &\qw \\
& \ctrl{-1}  \qw& \qw &&& 
\gate{H} & \targ & \gate{H} & \qw \\
}
\label{eq:CCX}
\end{equation}

While earlier studies have focused on optimizing the Toffoli gate at the pulse level~\citep{Hardware-Conscious_24}, these optimizations often introduce an unintended SWAP gate between the control qubits. Although this side effect can be useful in some cases, it may not always be desirable, depending on the specific application or hardware constraints.

A method for optimizing quantum circuits by considering the native gate set of the hardware is proposed in~\citep{Hardware-Conscious}. The approach, applied to the Toffoli gate on IBM Quantum, reduces gate infidelity by 18\% and cuts the required multi-qubit gates from eight to six by incorporating multi-qubit cross-resonance (MCR) gates. This optimization improves efficiency for linearly connected qubits~\citep{Hardware-Conscious}.

\subsection{Toffoli via ECR gates}

The impact of qubit connectivity on quantum algorithm performance is a crucial aspect of quantum computing hardware design. While many quantum algorithms have been developed with ideal all-to-all connectivity in mind, near-term practical quantum devices often feature restricted connectivity due to physical and engineering constraints~\citep{qubitconnectivity24}. This limitation can affect the performance and efficiency of quantum algorithms. However, by tailoring quantum algorithms to the specific connectivity constraints of the hardware, it is possible to significantly improve performance, making practical quantum computing more achievable even with less-connected qubit architectures~\citep{qubitconnectivity20}.

\begin{figure*}
\centering
\begin{equation}
\label{eq:CCX_ECR}
\text{\includegraphics[width=\textwidth]{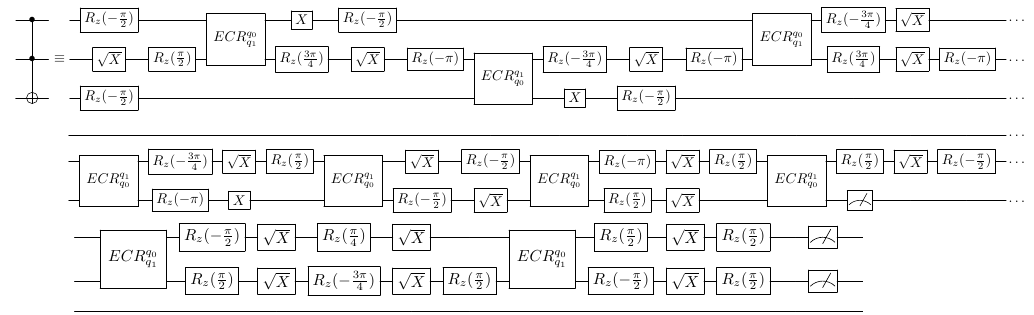}}
\end{equation}
\end{figure*}

For this purpose, we propose decomposing the CCX gate into nine ECR gates, as detailed in Eq.~\eqref{eq:CCX_ECR}. This method leverages the advanced capabilities of ECR gates to reduce circuit depth and enhance overall gate efficiency, addressing the limitations posed by qubit connectivity and improving practical implementation on NISQ devices.

In Eq.~\eqref{eq:CCX_ECR}, \(\sqrt{X}\) denotes the square root of the NOT gate, a gate that applies a half-bit-flip (half-X) operation. Additionally, \(R_z\) represents 1-qubit rotations about the \(z\)-axis. This quantum circuit can be further optimized to reduce the number of 1-qubit operations by exploiting symmetries or combining operations, leading to a more efficient implementation that minimizes gate depth and qubit usage. The matrix representations of the 2-qubit operations \(ECR^{q_0}_{q_1}\) and \(ECR^{q_1}_{q_0}\) are shown in Eq.~\eqref{eq:ecr1} and Eq.~\eqref{eq:ecr2}, respectively.

\begin{equation} \label{eq:ecr1}
ECR ^{q_0}_{q_1} = \frac{1}{\sqrt{2}} \begin{pmatrix} 0 & 1 & 0 & i \\ 1 & 0 & -i & 0 \\ 0 & i & 0 & 1 \\ -i & 0 & 1 & 0 \end{pmatrix}
\end{equation}

\begin{equation} \label{eq:ecr2}
ECR ^{q_1}_{q_0} = \frac{1}{\sqrt{2}} \begin{pmatrix} 0 & 0 & 1 & i \\ 0 & 0 & i & 1 \\ 1 & -i & 0 & 0 \\ -i & 1 & 0 & 0 \end{pmatrix}
\end{equation}

\section{Conclusion}

As quantum computing advances towards practical utility~\citep{IBM-Berkeley}, addressing the limitations of noisy intermediate-scale quantum (NISQ) devices becomes increasingly critical~\citep{NISQ18}. These devices are constrained by limited qubit connectivity, which necessitates the implementation of long-range entangling gates in linear depth, complicating quantum algorithm execution. Additionally, the gradual decline in gate fidelity, likely due to device aging, further challenges entanglement capabilities. In this context, optimizing quantum circuits is essential to mitigate these issues~\citep{NoiseOptimization}. 

Among the various quantum gates, the multi-qubit Toffoli gates stand out due to their critical role in quantum error correction and its application in a wide array of complex quantum algorithms. Despite its significance, the practical deployment of the Toffoli gate (on NISQ devices) is hindered by the challenges of circuit depth, gate infidelity, and the inherent complexity of its multi-qubit nature~\citep{Toffoli}.

This paper proposes a novel approach for decomposing the 3-qubit Toffoli gate into Echoed Cross-Resonance (ECR) gates. The ECR gate~\citep{cross-resonance, ECR_Hamiltonian, 64QV, ibm433} is a crucial quantum gate designed to efficiently generate entanglement between qubits~\citep{Entangling,SQSCZ2}. This gate is maximally entangling and can be considered equivalent to a CNOT gate when accounting for 1-qubit pre-rotations. Unlike the CNOT gate, the ECR gate uses an echoing procedure to mitigate undesired interactions, specifically those not of the ZX form, by canceling them out during gate operation~\citep{ECR_Hamiltonian, 64QV}.

The primary objective of the ECR gate is to enhance the fidelity of 2-qubit operations by reducing specific types of errors, thereby achieving more robust quantum gates. The key mechanism behind this improvement is the principle of \textit{echoing}, where the gate is applied in a manner that counters errors by adjusting the timing and parameters of the operations.

Unlike traditional decomposition methods, which often assume ideal qubit connectivity, our approach takes advantage of the ECR gate's native alignment with superconducting qubit systems. This reduces the complexity and depth typically required for Toffoli gate implementations on NISQ devices~\citep{qubitconnectivity24}. Not only does this approach simplify the gate's realization on NISQ devices, but it also provides a pathway for enhancing gate fidelity and minimizing error accumulation, addressing the critical need for optimizing quantum operations on near-term devices.

\section*{Declarations}

\subsection*{Consent for publication}
``The author have approved the publication. This research did not involve any human, animal or other participants."

\subsection*{Availability of supporting data}
``The datasets generated during and/or analyzed during the current study are included within this article."

\subsection*{Competing interests}
``The author declares no competing interests."

\subsection*{Funding}

``The author declares that no funding, grants, or other forms of support were received at any point throughout this research."

\end{document}